\documentclass[a4paper,longbibliography,twocolumn,amsmath,amssymb,amsfonts,superscriptaddress,floatfix,showpacs]{revtex4-1}
\usepackage[utf8]{inputenc}
\usepackage[english]{babel}
\usepackage[colorlinks=true,citecolor=green,linkcolor=red,urlcolor=blue]{hyperref}
\usepackage{braket}
\usepackage{bbm}
\usepackage{graphicx}
\usepackage{color}
\usepackage{caption}
\usepackage{subcaption}
\graphicspath{{./figs/}}%
\usepackage[normalem]{ulem}

\newcommand{\skippart}[1]{}

\newcommand{\be}{\begin{equation}}
\newcommand{\ee}{\end{equation}}
\newcommand{\bea}{\begin{eqnarray}}
\newcommand{\eea}{\end{eqnarray}}

\begin{document}

\title{Entanglement compression in scale space: \\
from the multiscale entanglement renormalization ansatz to matrix product operators}
\author{Karel~\surname{Van Acoleyen}}
\author{Andrew~\surname{Hallam}}
\author{Matthias~\surname{Bal}}
\author{Markus~\surname{Hauru}}
\author{Jutho~\surname{Haegeman}}
\author{Frank~\surname{Verstraete}}
\affiliation{Department of Physics and Astronomy, Ghent University, Krijgslaan 281, S9, B-9000 Ghent, Belgium}

\begin{abstract}
The multiscale entanglement renormalization ansatz (MERA) provides a constructive algorithm for realizing wavefunctions that are inherently scale invariant. Unlike conformally invariant partition functions however, the finite bond dimension $\chi$ of the MERA provides a cut-off in the fields that can be realized. In this letter, we demonstrate that this cut-off is equivalent to the one obtained when approximating a thermal state of a critical Hamiltonian with a matrix product operator (MPO) of finite bond dimension $\chi$. This is achieved by constructing an explicit mapping between the isometries of a MERA and the local tensors of the MPO.
\end{abstract}
\maketitle

\noindent \emph{Introduction.} Arguably the central problem in the study of extended quantum many-body systems - whether it is in the context of high energy physics or condensed matter physics - is the identification of the relevant infrared (IR) degrees of freedom. A new perspective on this problem arose with the advent of tensor network states (TNS) \cite{Cirac_2009,Verstraete2008}. The crucial feature of the TNS approach is that it directly targets the quantum many-body state and in particular its entanglement structure. In this sense TNS realise a compression or truncation of the exact quantum state in terms of the most relevant entanglement degrees of freedom. In particular, as the virtual (contracted) indices of the tensors encode the entanglement structure, the bond dimension $\chi$ of these indices can be thought of as the systematic control parameter of the \emph{entanglement compression} achieved by a particular TNS. Given their success thus far it is desirable to better understand the precise nature of this compression. In this letter we focus on systems with one spatial dimension.      
 
For matrix product states (MPS), this picture of entanglement compression was recently made more concrete in terms of the Euclidean path integral representation of the state  \cite{Zauner_2015,Rams_2015,Bal_2016}. In particular it was argued that MPS approximations can be understood as efficient compressions of the strip path integral over a finite imaginary time interval $L$, with $L$ scaling polynomially in $\chi$. This explains why MPS approximations for ground-states of gapped systems are essentially exact for sufficiently large bond dimension, while for critical systems this motivates the use of finite entanglement scaling methods in direct correspondence with the finite length scaling methods of Monte-Carlo simulations \cite{nishino1996numerical,Pollmann_2008,Tagliacozzo2008,Pirvu_2012,vanhecke2019scaling}.  

In this letter we address the role of the finite bond dimension and the type of compression in the case of the multi-scale entanglement renormalization ansatz (MERA) \cite{Vidal_2007,vidal_class_2008,Evenbly_2009}. This is a different type of TNS,  that is tailored specifically towards the description of critical systems. A remarkable feature of this ansatz is that it realises an explicit renormalization group (RG) flow directly at the level of the quantum state: the network has an extra \emph{scale} dimension with subsequent layers of local isometries and unitaries (disentanglers) effecting the RG flow. In contrast to MPS, this structure allows to capture the power-law decay of correlations and the logarithmic growth of the finite interval entanglement entropy, already at finite bond dimension. Furthermore, the conformal data of the theory - the different scaling dimensions and operator product expansion (OPE) coefficients - can be extracted directly from local scaling operators that are constructed in terms of the MERA isometries and unitaries \cite{Giovannetti_2008,Pfeifer_2009,evenbly2011quantum}. Notice that it has been shown already that one can construct MERA states directly from the application of the tensor network renormalization (TNR) procedure on the half-finite Euclidean path integral with open physical indices  \cite{Evenbly_2015}. But regarding the role of the bond dimension this merely shifts the question to the role of the bond dimension in TNR. 
 
Here we will offer a different view: we will argue that MERA approximations for critical systems can be understood naturally in \emph{scale space}. For this we revisit and further develop the results of \cite{czech_tensor_2016} on the MERA realisation of the conformal plane to strip map, which for the MERA is a map from physical to scale space; see also \cite{Evenbly_2010,Evenbly_2013} for related approaches. In particular we will argue and show numerically that the half-infinite subsystem density matrix reduces to a thermal (strip) state in scale space, $\rho_{scale}=e^{-\beta \bar{H}}$, for a Hamiltonian $\bar{H}$ belonging to the same universality class as the original critical Hamiltonian $H$ for which the MERA was optimised. The picture that is emerging is one where a finite $\chi$ MERA yields a particularly efficient matrix product operator (MPO) approximation for thermal states. Intriguingly, our numerics indicate that $\chi$ is controlling both an IR and ultraviolet (UV) cut-off of the CFT Hamiltonian $\bar{H}$.  Furthermore, the (single-site) MERA scaling operator appears as the transfer matrix of this strip state, which relates the MERA procedure for extracting scaling dimensions to Cardy's finite-size scaling \cite{Cardy_1984,Cardy:1986ie,Cardy1988}. Finally, motivated by the MPO compression view, in the last part of this letter we propose and implement a new algorithm for the construction of isometries and corresponding scaling operators from direct MPO approximations of the thermal state in terms of the original Hamiltonian $H$. \vspace{0.1cm}  

\noindent \emph{From real space to scale space, from the plane to the strip.} It is useful to start with a brief review of some CFT results in the continuum as this will guide our interpretation of the discretised MERA states. We take a CFT on the infinite line and consider the vacuum density matrix for the infinite right half-line subsystem. First, for any relativistic QFT (including CFTs) a foliation of the Euclidean path integral in the angular direction leads to the general expression (see e.g. \cite{Witten_2018}): \be \rho=\exp^{-2\pi \int^{+\infty}_{0} \!\!dx\, x\, T_{00}(x)}\,,\ee where $T_{00}(x)$ is the energy density operator. Here and in the following we drop the normalisation constant ensuring $Tr \rho=1$. For CFTs then, the conformal map from the plane to the strip with width $\beta$, \be z\rightarrow w= \frac{\beta}{2\pi} \log z\,,\ee maps this density matrix to a Gibbs state that is translation invariant: \be \rho_{strip}=\exp^{-\beta \int_{-\infty}^{+\infty}\!\!ds\,\, \bar{T}_{00}(s)}\,.\ee  Here $z=x+it$ and $w=s+i\tau$. Notice that at the level of Hilbert spaces, this conformal map boils down to a canonical (isometric) map $W$ of the original CFT Hilbert space on the half-line subsystem $x\in[0,+\infty ]$ to a new CFT Hilbert space on the full line $s\in [-\infty,+\infty]$, with e.g. the transformation for the energy density operator:  $W^{+}T_{00}(x)W=(\frac{ds}{dx})^{2} \bar{T}_{00}(s)=(\frac{\beta}{2\pi x})^2 \bar{T}_{00}(s)$, where $\bar{T}_{00}(s)$ is the energy density operator on the new Hilbert space and we drop the Schwarzian derivative term proportional to the unit operator \cite{DiFrancesco:639405}. It is readily checked then that this map $W$ indeed transforms $\rho$ into $\rho_{strip}$: $W^+\rho\, W=\rho_{strip}$. 

\begin{figure}[t]
\begin{subfigure}[b]{.5\textwidth}
\includegraphics[width=\textwidth]{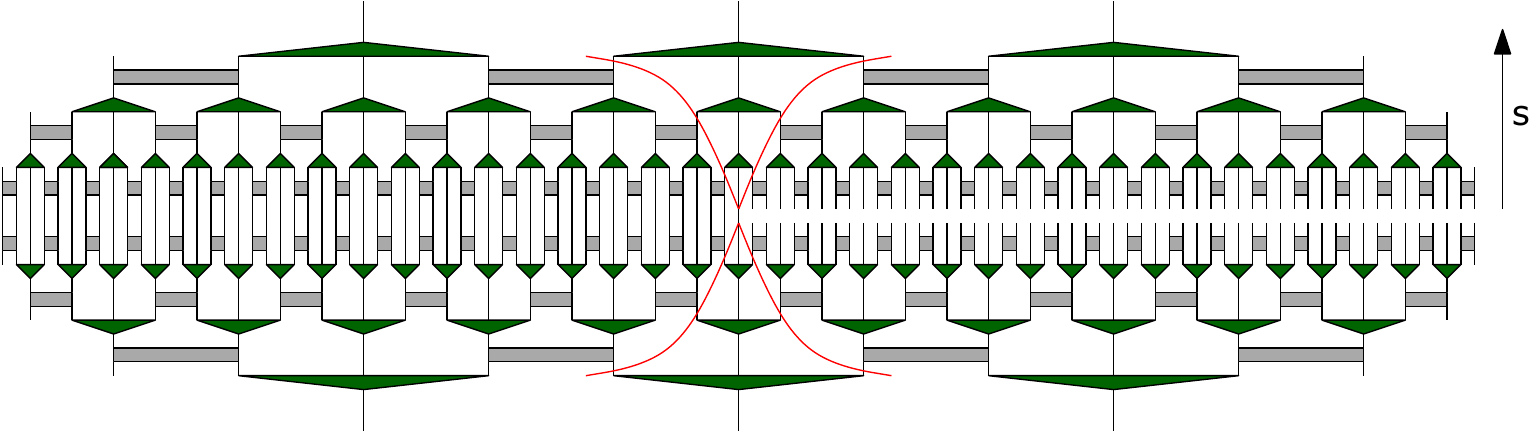}
\caption{\label{fig:ternarymera}}
\end{subfigure}
\begin{subfigure}[b]{.24\textwidth}
\includegraphics[width=.3\textwidth]{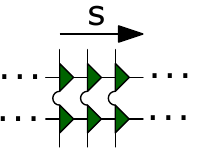}
\caption{\label{fig:mpo}}
\end{subfigure}\hfill
\begin{subfigure}[b]{.24\textwidth}
\includegraphics[width=.3\textwidth]{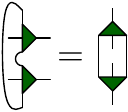}
\caption{\label{fig:scalingoperator}}
\end{subfigure}
\captionsetup{justification=raggedright}
\caption{(a): The MERA expression for the reduced density matrix on the right infinite half-line. (b): The MPO expression for the reduced density matrix in scale space $\rho_{scale}$, constructed out of the isometries and their conjugates. (c): The corresponding transfer operator for the partition function that is equal to the single-site MERA scaling operator.  }
\end{figure}

We now consider the discretised version of a half-line CFT density matrix as realised by a numerically optimised ternary MERA with bond dimension $\chi$. As illustrated in fig.\ref{fig:ternarymera}, by contracting the ket and bra to the left of the origin, we obtain the reduced density matrix $\rho$ defined on the real space interval $x \in~[0,+\infty]$. All tensors to the left of the causal cone of the origin site drop out due to the unitary and isometric restrictions on the tensors, resulting in an explicit expression $\rho=W \rho_{scale} W^+$. Here $\rho_{scale}$ is the matrix product operator (MPO) of fig.\ref{fig:mpo}, with the MPO tensors constructed out of the MERA isometries and their conjugates at the different layers labeled by the RG log-scale $s$. $W$ is the full isometry from the physical real space indices on the right half-line, to the virtual indices in scale space that are cut by the right causal cone. For a critical MERA $\rho_{scale}$ will be translation invariant up to the first few $s$-dependent isometries that capture the non-universal UV behaviour of the state. The previous continuum CFT discussion then strongly suggests that $W$ gives us a discretised version of the conformal plane to strip map, in particular with $z=3^w$, as local operators at position $x$ in physical space get mapped to local operators at position $s\sim \log_3 (x)$ in scale space. Notice that different from \cite{czech_tensor_2016}, we consider here both the bra and ket, and as a consequence the map $W$ is exactly isometric.

Taking $W$ as the discretized plane to strip map we get the prediction that the bulk density matrix in scale space is a thermal (strip) density matrix with inverse temperature $\beta=\frac{2\pi}{\log(3)}$ of a Hamiltonian $\bar{H}$ with the same IR CFT fixed point as the original Hamiltonian for which the MERA was optimised: \be \rho_{scale}=\exp^{-\frac{2\pi}{\log (3)} \bar{H}}\,. \label{prediction} \ee
Before discussing the numerics in light of this prediction, let us point out that this ties in nicely with Cardy's seminal results on finite-size scaling for CFTs. From the conformal plane to strip map it was shown that the correlation functions on the strip should decay exponentially, in particular the transfer matrix $T$ on the strip (with periodic boundary conditions) should have the spectrum \be T=\exp^{-\frac{2\pi}{\beta}\Delta}\,,\label{finitesizescaling}\ee with $\Delta$ the scaling dimensions of the local primary operators and their descendants \cite{Cardy_1984,Cardy:1986ie,Cardy1988}. As shown explicitly in fig.\ref{fig:scalingoperator}, upon a proper identification of the indices the transfer matrix of $\rho_{scale}$ indeed precisely reduces to the single-site MERA scaling (super-)operator $S$ that is used to extract the scaling dimensions from the MERA simulation according to $S=\exp^{-\log(3) \Delta}$.

\begin{figure}[t]
\begin{subfigure}[b]{.23\textwidth}
\includegraphics[width=\textwidth]{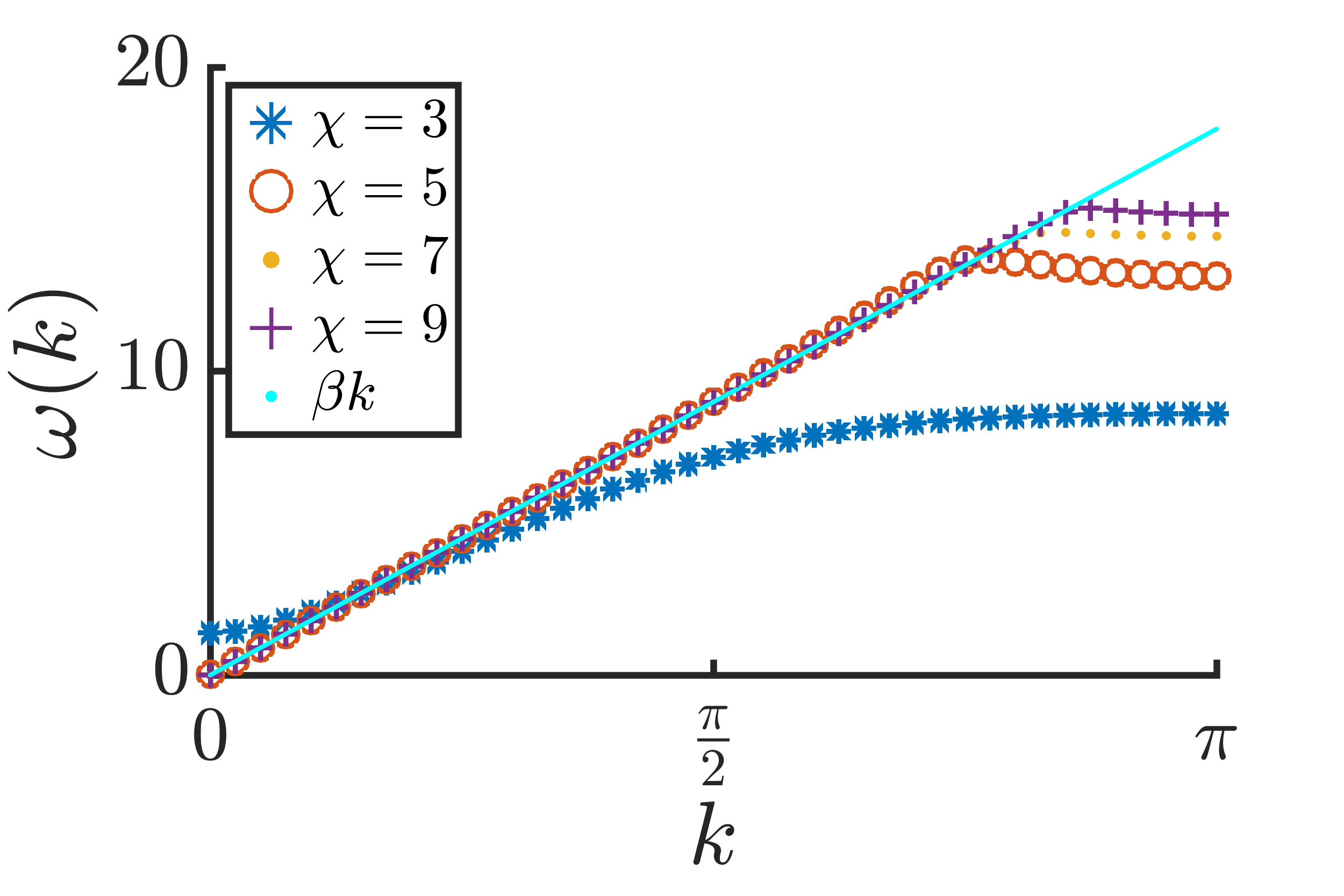}
\caption{\label{fig:isingdispersion1}}
\end{subfigure}
\begin{subfigure}[b]{.23\textwidth}
\includegraphics[width=\textwidth]{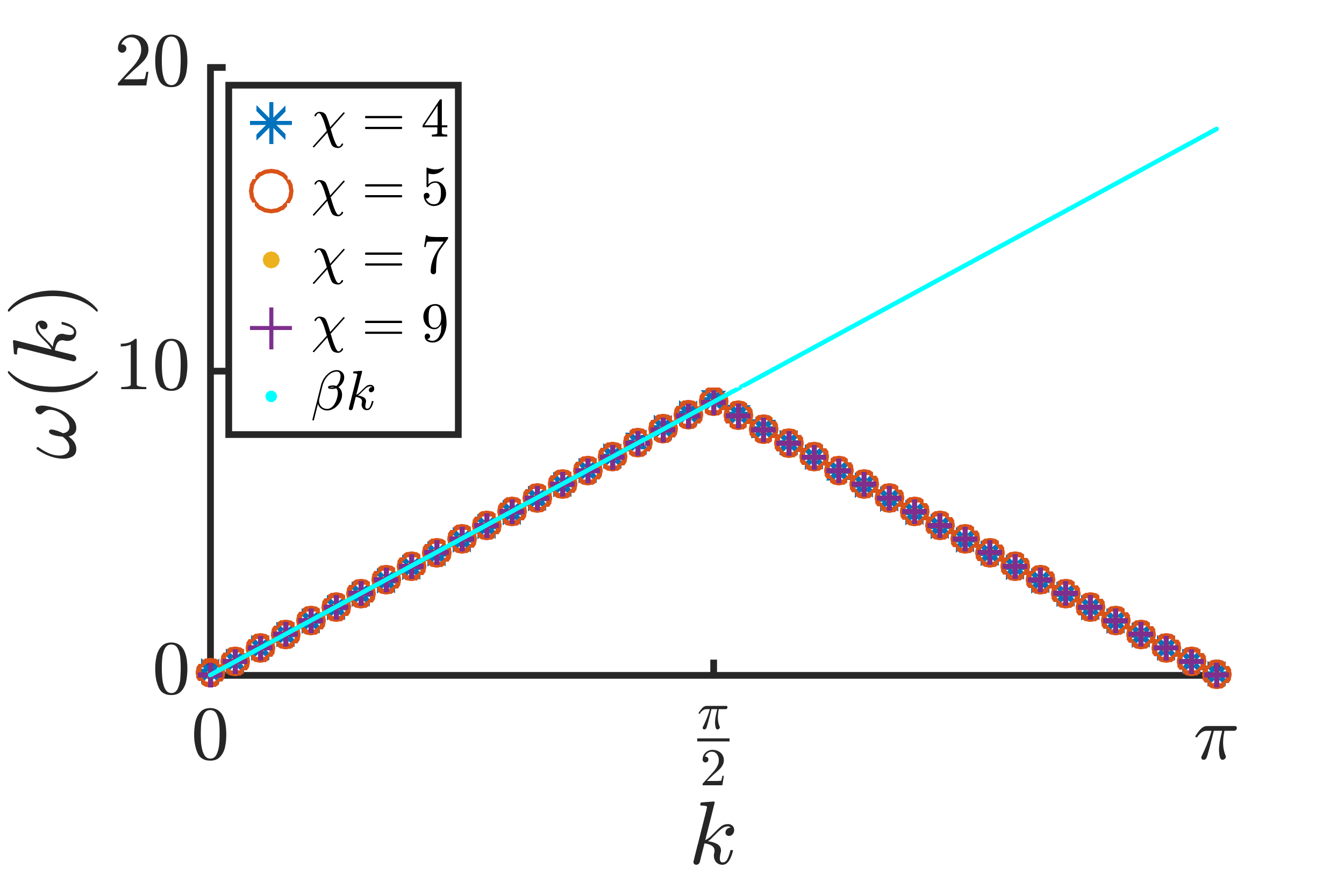}
\caption{\label{fig:isingdispersion2}}
\end{subfigure}\hfill
\captionsetup{justification=raggedright}
\caption{(a): The log spectrum of the MPO $\rho_{scale}$ (fig.\ref{fig:mpo}), $\omega=-\log \rho_{scale}$, in the thermodynamic limit, for the bulk isometries optimised on the critical Ising model, for different bond dimensions $\chi$. (b): The comparable log spectrum for bulk isometries initialized differently (see text). In each case the blue solid line shows the linear dispersion with gradient $\beta=\frac{2\pi}{\log{3}}$. }
\label{fig:isingdispersion}
\end{figure}

We have studied the properties of $\rho_{scale}$  at different MERA bond dimensions $\chi$, for the transverse-field Ising model at its critical point. First we have numerically optimised all the MERA tensors with the standard MERA algorithm \cite{Evenbly_2009} - including two non-scale invariant UV layers, imposing the explicit $\mathbb{Z}_2$ symmetry of the Ising model \cite{singh_tensor_2011,singh_symmetry_2013}. Then we have considered the MPO expression for $\rho_{scale}$ constructed out of the (scale-independent) bulk isometries (fig. \ref{fig:mpo}). Using standard matrix product state (MPS) techniques, we then have constructed an MPS approximation of the fixed point of the MPO in the thermodynamic limit, and have used the MPS particle momentum eigenstate ansatz \cite{Haegeman_2012}, to approximate the one-particle excitations on top of the fixed point. 

In fig.\ref{fig:isingdispersion} we show the resulting log spectrum, $\omega=-\log \rho_{scale}$ for MERA optimizations on the Ising model for different $\chi$. Notice a subtlety involving a MERA gauge choice: depending on this choice one ends up with an isometry that gives rise to a positive or non-positive scaling operator $S$, and a dispersion relation of the type of fig.\ref{fig:isingdispersion1}  or fig.\ref{fig:isingdispersion2} \footnote{In our simulations, depending on the initial random seed in the numerical optimization we end up with isometries giving rise to two different types of dispersion relation: of the ferro-magnetic type fig.2a or of the anti-ferro-magnetic type fig.2b, with only a positive scaling operator $S$ in the former case.  By applying a $\mathbb{Z}_2$ transformation on the upper leg of the isometry we can switch between both cases. Using the $\mathbb{Z}_2$ symmetry of the tensors, one can show that this is a MERA gauge transformation, i.e. a transformation on the MERA tensors that leaves the actual physical state invariant. From the same $\mathbb{Z}_2$ symmetry of the tensors one can also show that this transformation shifts the momentum of the MPO eigenstates: $k\rightarrow k+m\pi$, with $m=(0,1)$ the $\mathbb{Z}_2$ quantum number of the state. This then explains the $k>\pi/2$ part of the second case fig.2b as arising from the $m=1$ negative momentum branch of the first case. This generalises to more general group symmetries: contracting an element of the group $U_g$ - with $g$ in the center of the group - on the upper leg of the symmetric isometry, shifts the momentum of the MPO eigenstates accordingly, while leaving the physical MERA state invariant.}. 

\begin{figure}[t]
\begin{subfigure}[b]{.23\textwidth}
\includegraphics[width=\textwidth]{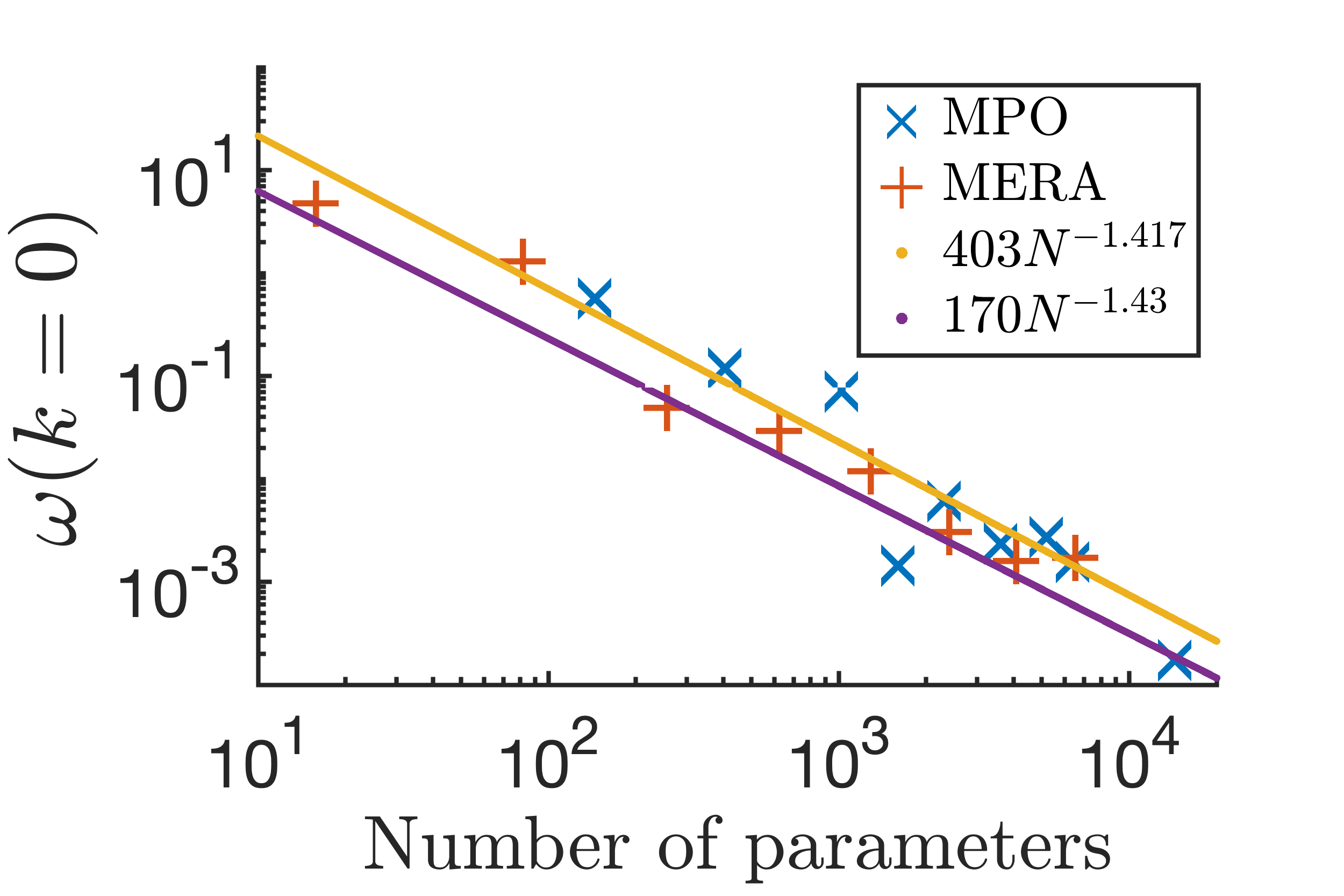}
\caption{\label{fig:gap}}
\end{subfigure}
\begin{subfigure}[b]{.23\textwidth}
\includegraphics[width=\textwidth]{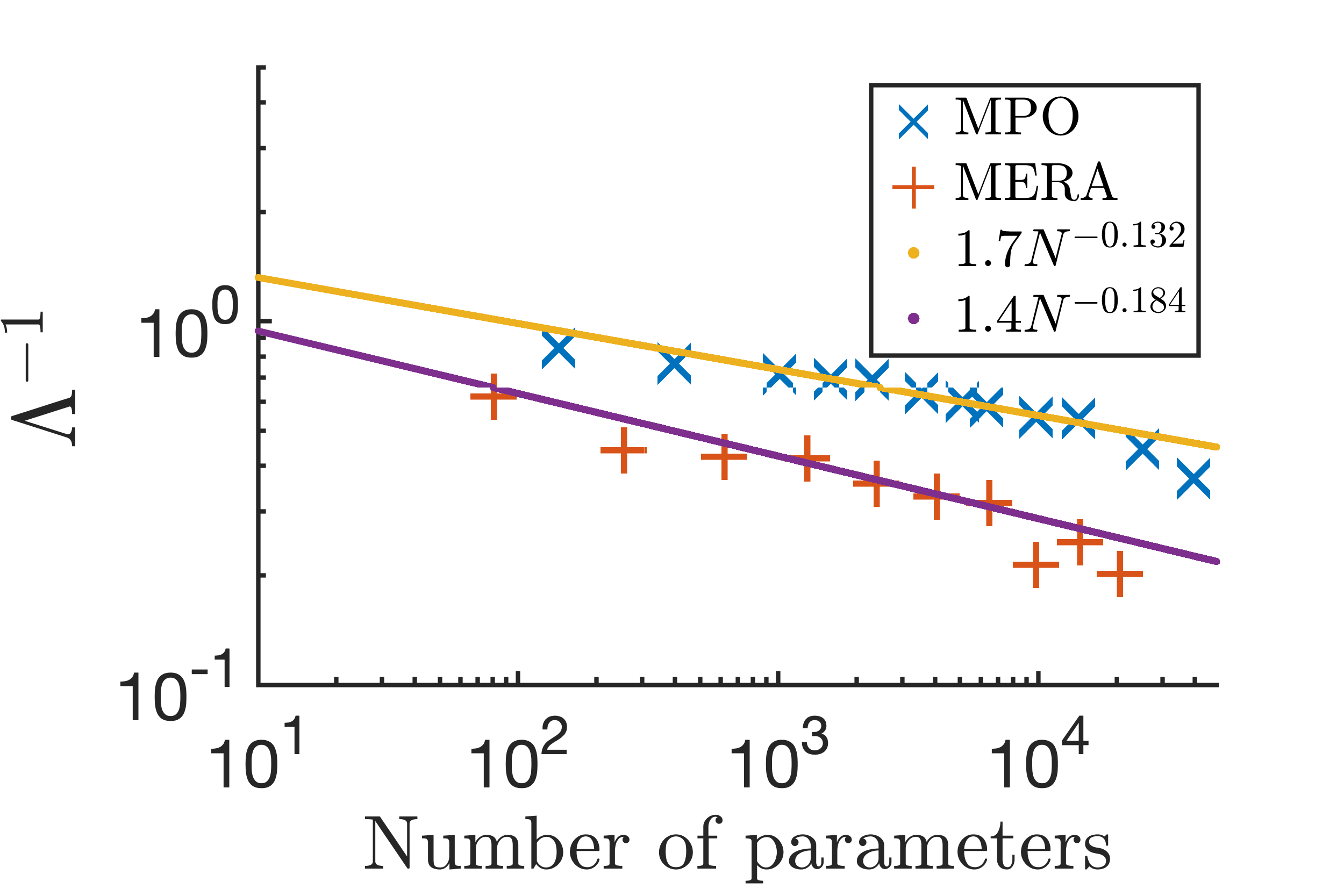}
\caption{\label{fig:UVcutoff}}
\end{subfigure}
\begin{subfigure}[b]{.23\textwidth}
\includegraphics[width=\textwidth]{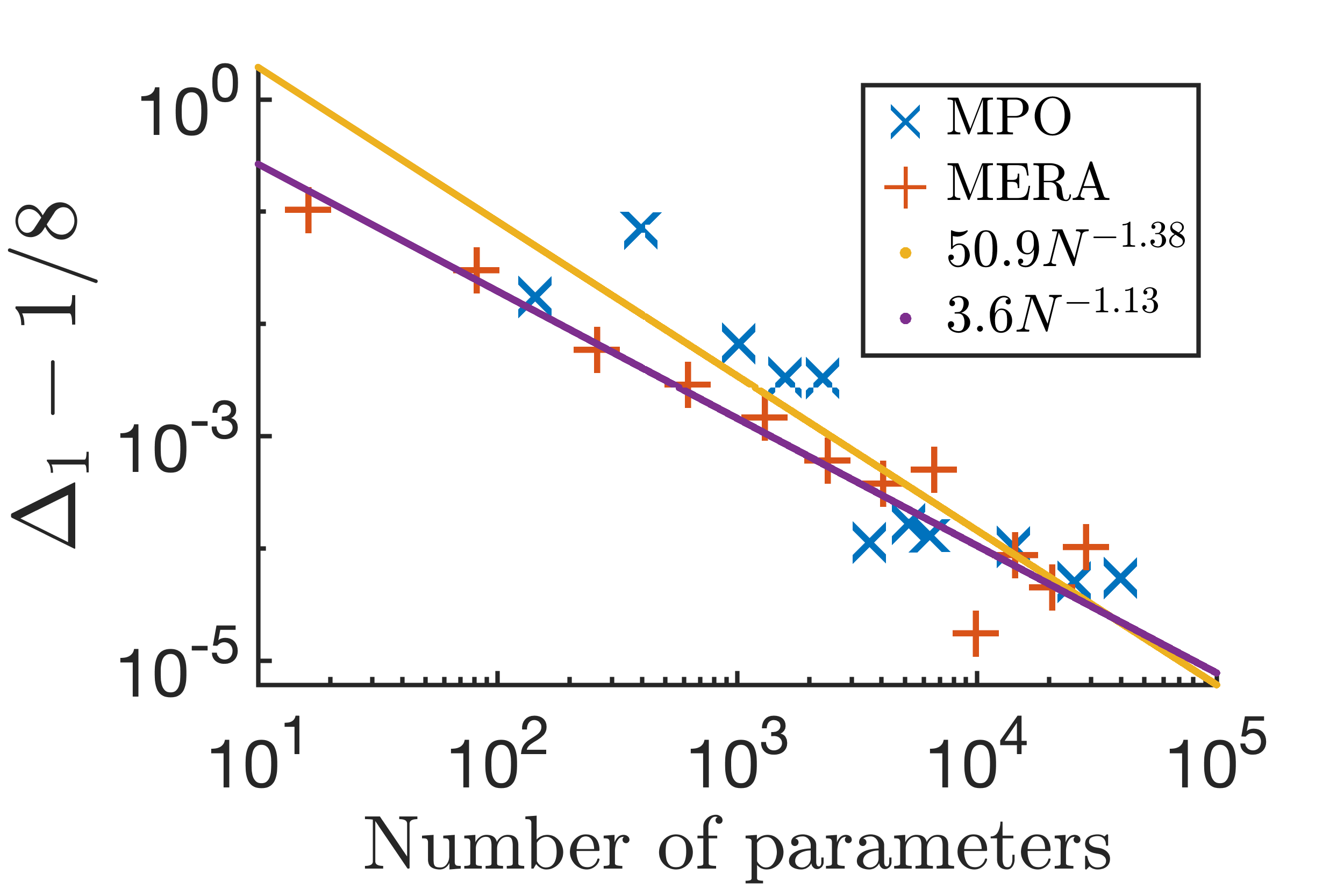}
\caption{\label{fig:error}}
\end{subfigure}
\begin{subfigure}[b]{.23\textwidth}
\includegraphics[width=\textwidth]{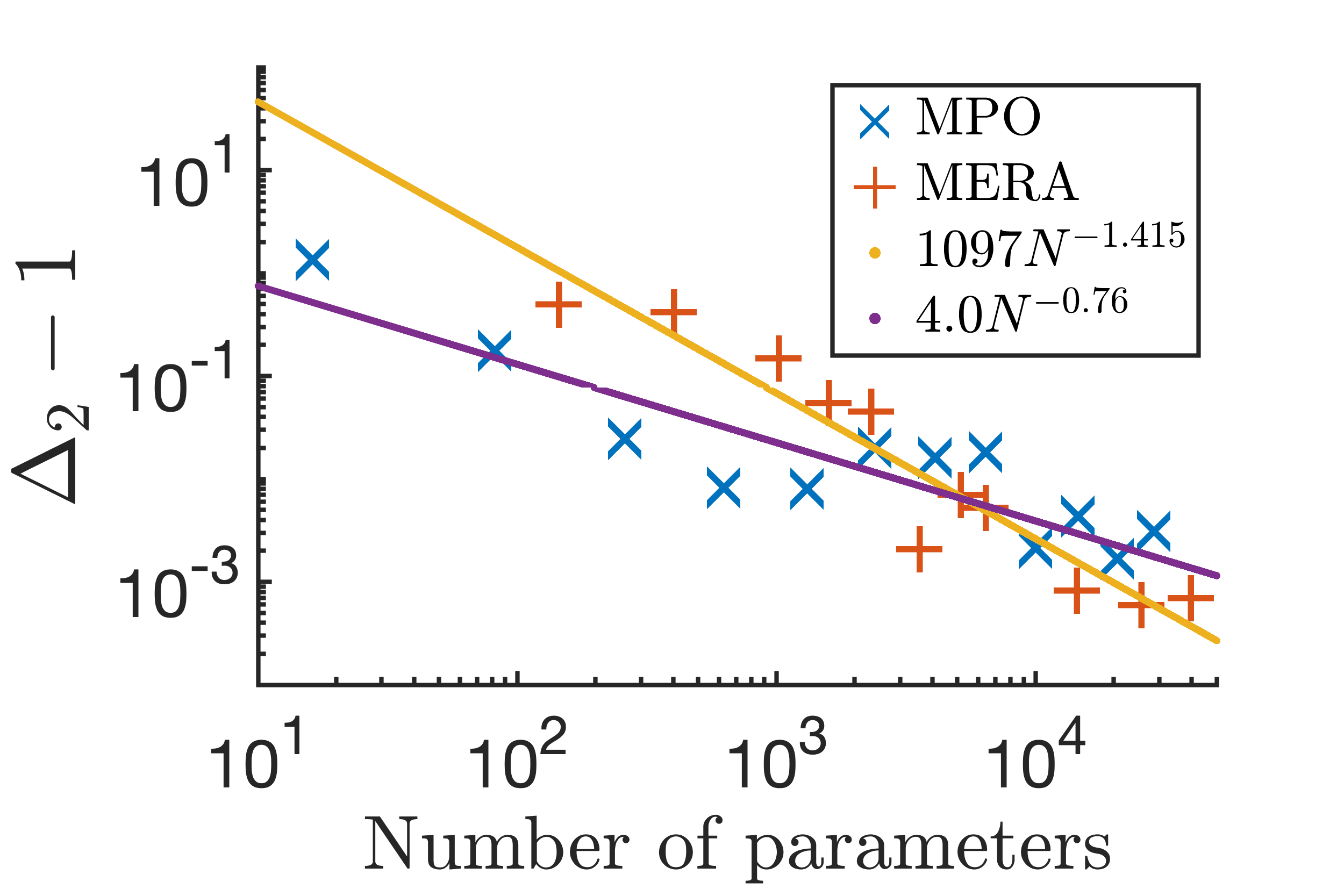}
\caption{\label{fig:error2}}
\end{subfigure}

\captionsetup{justification=raggedright}
\caption{For our Ising model simulations, with isometries obtained from MERA optimizations (red) and from direct MPO compressions (blue). (a): The IR gap in the dispersion relation, (b): The inverse UV cutoff, $\Lambda^{-1}$ in the dispersion relation. (c),(d): The error in the first two  scaling dimensions.  All as a function of the number of variational parameters $N$, with $N=\chi^4$ in the MERA case and $N=16D^2$ in the direct MPO case. }
\label{fig:errorscaling}
\end{figure}

From the prediction (\ref{prediction}) we would expect a momentum regime with a linear spectrum, $\omega(k)=\beta k$, with $\beta=2\pi/log(3)$. For the different $\chi$ values we indeed find a linear regime in very good agreement with this prediction. To discuss the role of $\chi$ we focus on the dispersion relations of the first type (fig.\ref{fig:isingdispersion1}). It turns out there is a small mass gap for $k\rightarrow 0$, which for growing $\chi$ decreases as an approximate power law, as can be seen in fig.\ref{fig:gap} (red crosses).  Notice that we have made sure that this is not an artefact of our MPS approximation of the fixed point: our results were well converged in the MPS bond dimension. Furthermore $\chi$ also has its effect at the other side of the spectrum, where we again see a deviation from the linear result. To associate a UV energy scale $\Lambda$ to this effect, we have identified $\omega({\pi})=\beta \Lambda \sin (\pi/\Lambda)$. Notice that there does not seem to be a unique preferred way to identify this UV scale. The rationale for our particular choice is that the right-hand side corresponds to an Ising dispersion relation on a coarse grained lattice with $\mathcal{O}(\Lambda)$ sites blocked into one effective site. In fig.\ref{fig:UVcutoff} we see that $1/\Lambda$ again appears to obey an approximate power law in $\chi$. So, taken at face value the bond dimension $\chi$ seems to control both a relevant (IR) and irrelevant (UV) perturbation to the CFT. In the proper scaling regime both type of operators give power corrections to the finite size-scaling (\ref{finitesizescaling}) \cite{Cardy:1986ie,Henkel_1987,Reinicke_1987} which in our case should therefore translate to power corrections for the scaling dimensions obtained from the MERA scaling operator $S$. This is reasonably consistent with our numerical results, see fig.\ref{fig:error} and fig.\ref{fig:error2}.  \vspace{0.1cm}  

\begin{figure}[t]
\begin{subfigure}[b]{.23\textwidth}
\includegraphics[width=\textwidth]{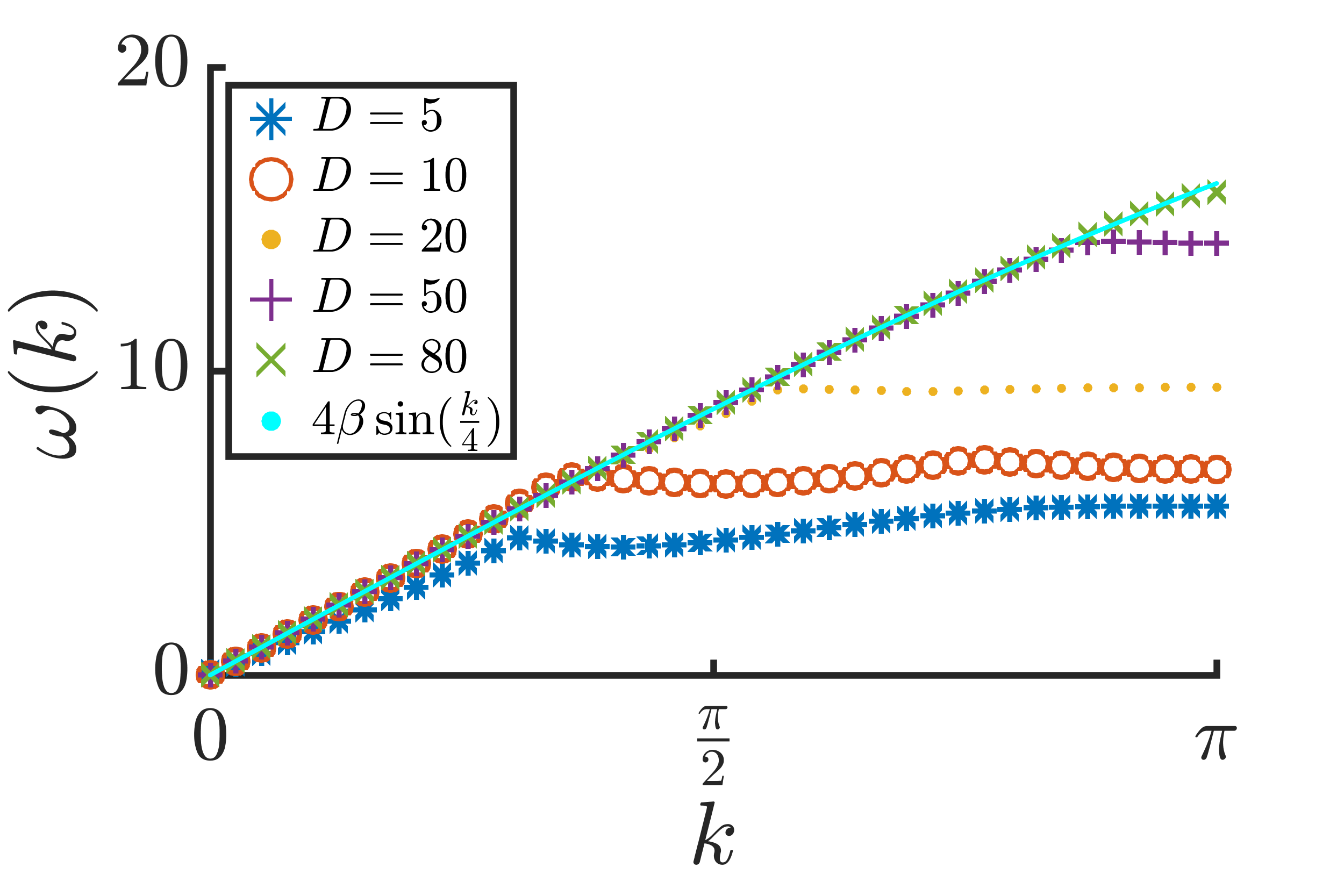}
\caption{\label{fig:IsingMPOdispersion}}
\end{subfigure}
\begin{subfigure}[b]{.23\textwidth}
\includegraphics[width=\textwidth]{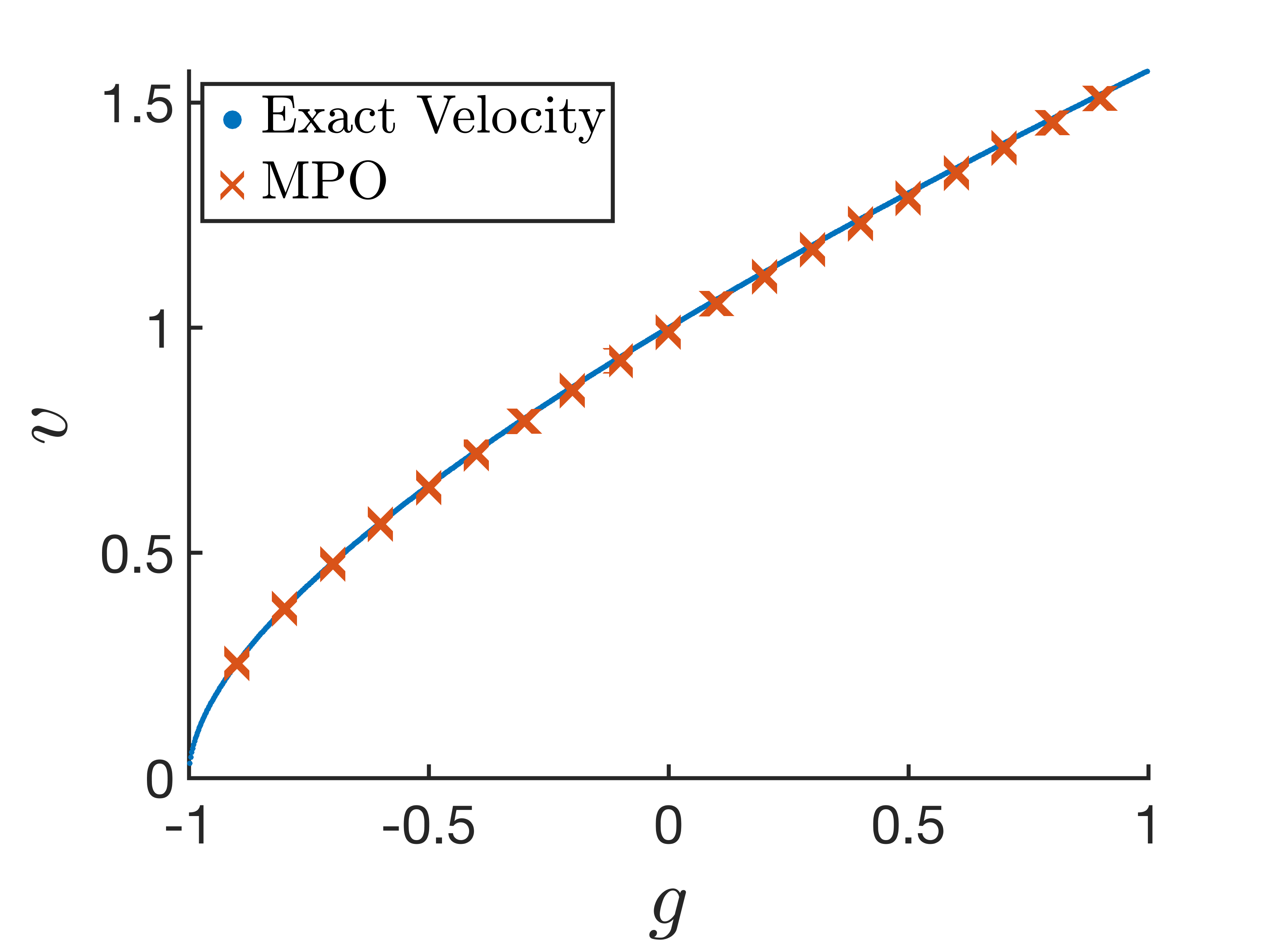}
\caption{\label{fig:XXZspeed}}
\end{subfigure}
\begin{subfigure}[b]{.23\textwidth}
\includegraphics[width=\textwidth]{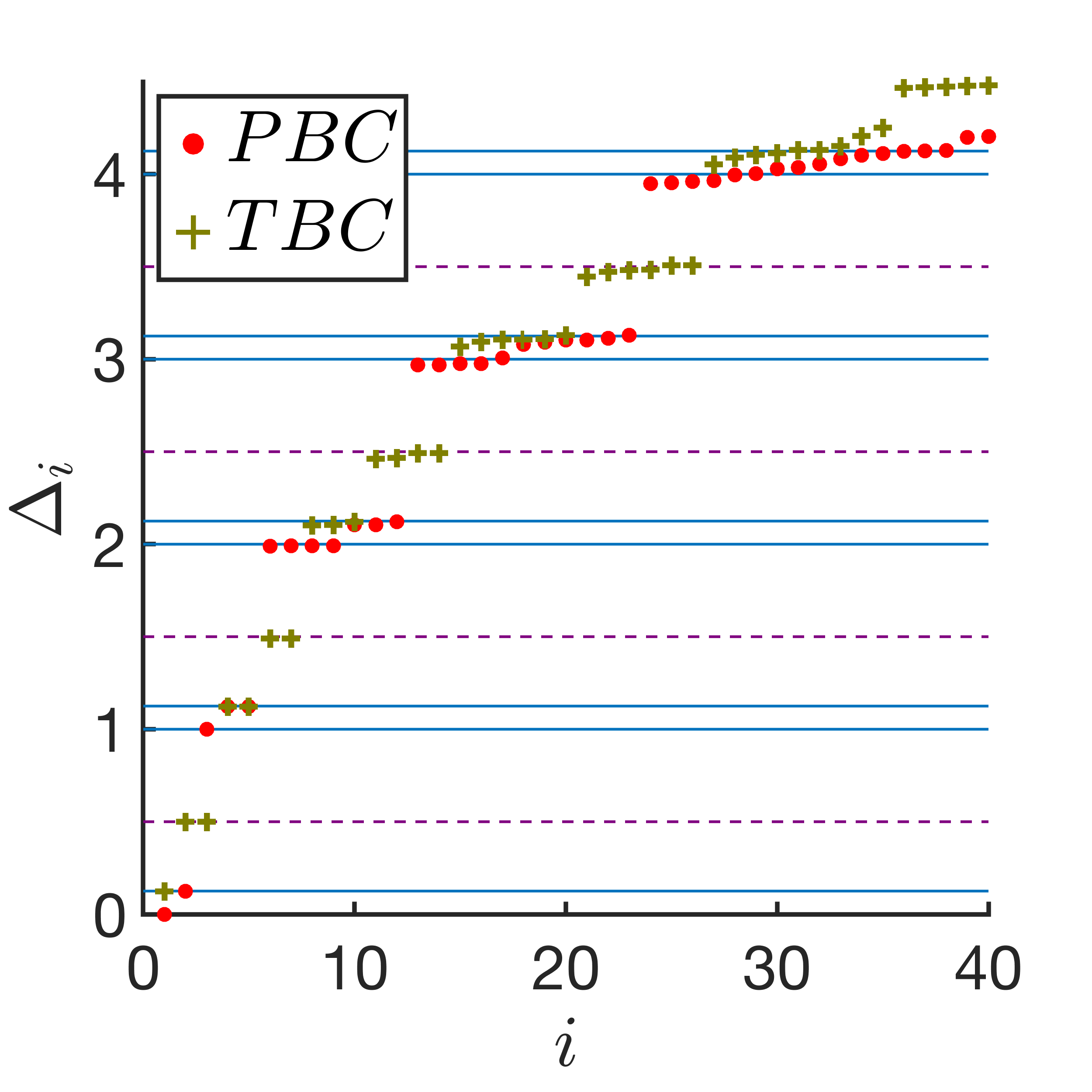}
\caption{\label{fig:Isingscaling}}
\end{subfigure}
\begin{subfigure}[b]{.23\textwidth}
\includegraphics[width=\textwidth]{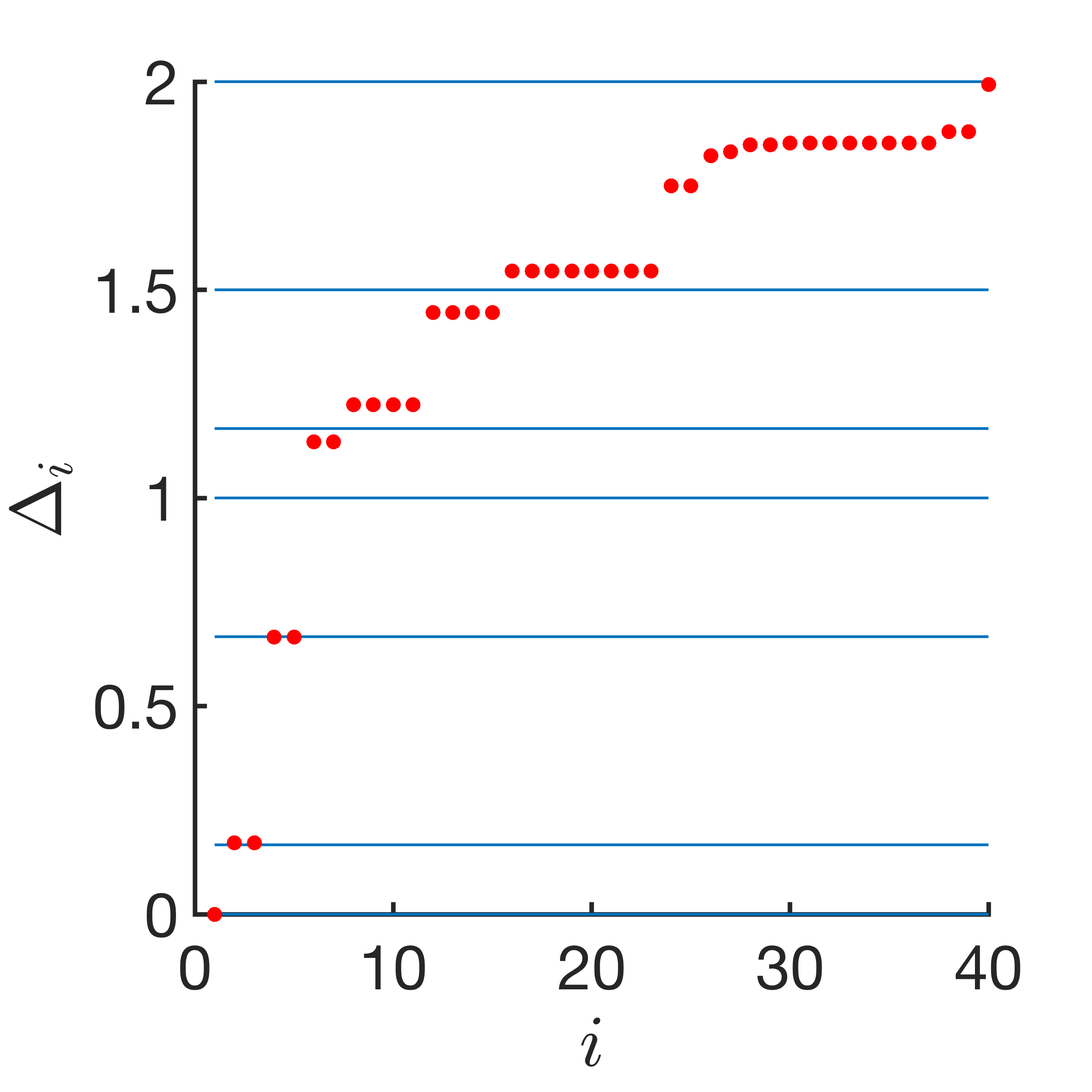}
\caption{\label{fig:XXZscaling}}
\end{subfigure}
\captionsetup{justification=raggedright}
\caption{(a): The log spectra of the MPOs $\omega=-\log \rho$, for MPO representations of the Ising thermal state at temperature $\beta=2\pi/\log{3}$ for different $D$. (b): The velocity of the free boson describing the XXZ model for different values of $g$ as calculated from the log-spectrum of the $D=30$ MPO, the solid line is the exact result \cite{Giamarchi}. (c): The Ising scaling dimensions from the MPO transfer operator for periodic (red circles) and twisted (green crosses) boundary conditions (at $D=80$).  (d): The XXZ (periodic boundary) scaling dimensions for $g=-1/2$ (at $D=200$). }
\label{fig:MPOresults}
\end{figure}

\noindent \emph{From thermal states to isometries and scaling operators.}
The results above suggest that a MERA optimisation is in effect performing an indirect compression of a thermal state for a critical Hamiltonian in terms of an MPO of the particular form fig.\ref{fig:mpo}. Our numerics indicate that this compression is efficient, with both the IR and UV cut-off scale, together with the corrections on the exact CFT scaling dimensions decaying according to a power law in $\chi$. As such this is no surprise as in general, both on theoretical grounds \cite{Hastings_2006,Molnar_2015} and based on many previous numerical applications, it is known that thermal states of local Hamiltonians can be efficiently approximated by an MPO. 

To make contact with the MERA results of the previous section we have considered a specific direct MPO compression scheme. In particular we have followed the approach of \cite{Verstraete_2004MPO}, by evolving the infinite-temperature state with the Hamiltonian in imaginary time using time-evolving block decimation in a second order Trotter approximation, with a timestep $\Delta t=\frac{\beta}{20000}$. In this way we have constructed an MPO approximation for $\rho=\exp^{-\beta H/2}$, for which the tensors in the proper (left-canonical) gauge are isometric. Upon multiplication of this MPO with its hermitian conjugate we arrive at an MPO approximation for $\rho=\exp^{-\beta H}$  of the specific form fig.\ref{fig:mpo}. 

Notice that the resulting isometries are not of the exact MERA form: the two outer lower legs of the isometry have a different bond dimension than the upper leg and center lower leg. The outer legs have a fixed bond dimension $d^2=4$ due to the blocking of two lattice sites, each with local Hilbert space dimension $d=2$. The other two legs have a bond dimension $D$ which controls the level of compression. By blocking more physical sites one could also generate MERA isometries with uniform bond dimension. Another obvious difference is that we now directly target the thermal state of the original microscopic Hamiltonian $H$, in contrast to the MERA case for which the entanglement Hamiltonian $\bar{H}$ only shares the IR properties with the Hamiltonian $H$ on which the MERA is optimised.          

In fig.\ref{fig:MPOresults} we show our results from this direct compression scheme for thermal states. Besides the critical Ising Hamiltonian we have also considered the $XXZ$ model. We were in principle free to chose $\beta$ but simply stayed with the previous value $\beta=2\pi/\log(3)$. Let us first discuss the Ising results. From the generated MPO approximations at different $D$ we obtained the log spectrum fig.\ref{fig:IsingMPOdispersion} in the exact same way as for the MPOs arising from the MERA simulations. The scaling dimensions of fig.\ref{fig:Isingscaling} were obtained from the transfer (scaling-) operator according to (\ref{finitesizescaling}). We also computed the non-local scaling dimensions by adding a twist $\sigma_z\otimes\sigma_z$ on one of the outer legs of the isometries, see \cite{Evenbly_2010nonlocal} for the same procedure on the MERA scaling operators. In fig.\ref{fig:errorscaling} one can see that the IR and UV scale for the direct MPO results (inferred from fig.\ref{fig:IsingMPOdispersion}) together with the errors on the scaling dimensions, indeed have a similar power law scaling behaviour as in the MERA case. Notice though that now $\Lambda$ should converge to $4$, in correspondence with the exact microscopic Ising result on our blocked lattice. For the MERA case, with growing local dimension $\chi$ we expect $\Lambda\rightarrow\infty$. 

We have performed simulations on the $XXZ$ Hamiltonian, $H=\sigma_x\sigma_x+\sigma_y\sigma_y-g \sigma_z\sigma_z$, along the full critical line $|g|<1$. Notice that for this model, in general the velocity $v(g) \neq 1$. This implies a linear dispersion $\omega(k)=\beta v(g) k/2$ on our (2 site) blocked lattice. From our results on the log spectrum of $\rho$ we can fit a value for $v(g)$ that agrees nicely with the exact result  \cite{Giamarchi}, as can be seen in fig.\ref{fig:XXZspeed}. The velocity $v(g)$ also shows up in the scaling dimension relation, which now reads: $S=\exp^{- \frac{4\pi}{\beta v(g)}\Delta}$. Fig.\ref{fig:XXZscaling} shows the $g=-0.5$ results from this scaling operator. The $XXZ$ model is significantly more challenging than the Ising model, this is reflected in the fact that a significantly higher bond dimension is needed to accurately determine its scaling dimensions. 

\vspace{0.1cm}

\noindent \emph{Outlook.}
Our results demonstrate that the finite bond dimension $\chi$ approximation of a MERA is of the same nature as the finite bond dimension $\chi$ MPO approximation of a thermal state of a critical Hamiltonian. This points to a new algorithm for creating critical MERA states. We have shown how to obtain the isometries, the complete algorithm of course also requires a similar construction for obtaining the disentanglers. But notice that the MPO expression of fig.\ref{fig:mpo} corresponds to a special cut. One can verify that arbitrary cuts lead to MPO expressions for $\rho_{scale}$ that also involve the disentanglers, giving rise to non-translation invariant transfer matrices that depending on $s$ take the form of the single-site scaling operator or the left, centre or right double-site scaling operators. This shows how to obtain the disentanglers: by minimising the breaking of translation invariance for MPOs corresponding to such general cuts \footnote{In preparation}. From the (bulk) disentanglers, in combination with the isometries, one can then also obtain the OPE coefficients, overcoming the main limitation of the thermal MPO approach that we presented here.  

 \vspace{0.1cm}

\noindent\emph{Acknowledgements.}
We would like to thank Laurens Vanderstraeten for inspiring discussions. This work is supported by an Odysseus grant from the FWO and has received funding from the European Research Council (ERC) under the European Union Horizon 2020 research and innovation programme (grant agreement No 647905 (QUTE), No 715861 (ERQUAF) and SIQS).

\bibliographystyle{apsrev4-1}
\bibliography{enthamscale}

\begin{thebibliography}{37}%
\makeatletter
\providecommand \@ifxundefined [1]{%
 \@ifx{#1\undefined}
}%
\providecommand \@ifnum [1]{%
 \ifnum #1\expandafter \@firstoftwo
 \else \expandafter \@secondoftwo
 \fi
}%
\providecommand \@ifx [1]{%
 \ifx #1\expandafter \@firstoftwo
 \else \expandafter \@secondoftwo
 \fi
}%
\providecommand \natexlab [1]{#1}%
\providecommand \enquote  [1]{``#1''}%
\providecommand \bibnamefont  [1]{#1}%
\providecommand \bibfnamefont [1]{#1}%
\providecommand \citenamefont [1]{#1}%
\providecommand \href@noop [0]{\@secondoftwo}%
\providecommand \href [0]{\begingroup \@sanitize@url \@href}%
\providecommand \@href[1]{\@@startlink{#1}\@@href}%
\providecommand \@@href[1]{\endgroup#1\@@endlink}%
\providecommand \@sanitize@url [0]{\catcode `\\12\catcode `\$12\catcode
  `\&12\catcode `\#12\catcode `\^12\catcode `\_12\catcode `\%12\relax}%
\providecommand \@@startlink[1]{}%
\providecommand \@@endlink[0]{}%
\providecommand \url  [0]{\begingroup\@sanitize@url \@url }%
\providecommand \@url [1]{\endgroup\@href {#1}{\urlprefix }}%
\providecommand \urlprefix  [0]{URL }%
\providecommand \Eprint [0]{\href }%
\providecommand \doibase [0]{http://dx.doi.org/}%
\providecommand \selectlanguage [0]{\@gobble}%
\providecommand \bibinfo  [0]{\@secondoftwo}%
\providecommand \bibfield  [0]{\@secondoftwo}%
\providecommand \translation [1]{[#1]}%
\providecommand \BibitemOpen [0]{}%
\providecommand \bibitemStop [0]{}%
\providecommand \bibitemNoStop [0]{.\EOS\space}%
\providecommand \EOS [0]{\spacefactor3000\relax}%
\providecommand \BibitemShut  [1]{\csname bibitem#1\endcsname}%
\let\auto@bib@innerbib\@empty
\bibitem [{\citenamefont {Cirac}\ and\ \citenamefont
  {Verstraete}(2009)}]{Cirac_2009}%
  \BibitemOpen
  \bibfield  {author} {\bibinfo {author} {\bibfnamefont {J.~I.}\ \bibnamefont
  {Cirac}}\ and\ \bibinfo {author} {\bibfnamefont {F.}~\bibnamefont
  {Verstraete}},\ }\href {\doibase 10.1088/1751-8113/42/50/504004} {\bibfield
  {journal} {\bibinfo  {journal} {Journal of Physics A: Mathematical and
  Theoretical}\ }\textbf {\bibinfo {volume} {42}},\ \bibinfo {pages} {504004}
  (\bibinfo {year} {2009})}\BibitemShut {NoStop}%
\bibitem [{\citenamefont {Verstraete}\ \emph {et~al.}(2008)\citenamefont
  {Verstraete}, \citenamefont {Murg},\ and\ \citenamefont
  {Cirac}}]{Verstraete2008}%
  \BibitemOpen
  \bibfield  {author} {\bibinfo {author} {\bibfnamefont {F.}~\bibnamefont
  {Verstraete}}, \bibinfo {author} {\bibfnamefont {V.}~\bibnamefont {Murg}}, \
  and\ \bibinfo {author} {\bibfnamefont {J.~I.}\ \bibnamefont {Cirac}},\ }\href
  {\doibase 10.1080/14789940801912366} {\bibfield  {journal} {\bibinfo
  {journal} {Advances in Physics}\ }\textbf {\bibinfo {volume} {57}},\ \bibinfo
  {pages} {143} (\bibinfo {year} {2008})}\BibitemShut {NoStop}%
\bibitem [{\citenamefont {{Zauner}}\ \emph {et~al.}(2015)\citenamefont
  {{Zauner}}, \citenamefont {{Draxler}}, \citenamefont {{Vanderstraeten}},
  \citenamefont {{Degroote}}, \citenamefont {{Haegeman}}, \citenamefont
  {{Rams}}, \citenamefont {{Stojevic}}, \citenamefont {{Schuch}},\ and\
  \citenamefont {{Verstraete}}}]{Zauner_2015}%
  \BibitemOpen
  \bibfield  {author} {\bibinfo {author} {\bibfnamefont {V.}~\bibnamefont
  {{Zauner}}}, \bibinfo {author} {\bibfnamefont {D.}~\bibnamefont {{Draxler}}},
  \bibinfo {author} {\bibfnamefont {L.}~\bibnamefont {{Vanderstraeten}}},
  \bibinfo {author} {\bibfnamefont {M.}~\bibnamefont {{Degroote}}}, \bibinfo
  {author} {\bibfnamefont {J.}~\bibnamefont {{Haegeman}}}, \bibinfo {author}
  {\bibfnamefont {M.~M.}\ \bibnamefont {{Rams}}}, \bibinfo {author}
  {\bibfnamefont {V.}~\bibnamefont {{Stojevic}}}, \bibinfo {author}
  {\bibfnamefont {N.}~\bibnamefont {{Schuch}}}, \ and\ \bibinfo {author}
  {\bibfnamefont {F.}~\bibnamefont {{Verstraete}}},\ }\href {\doibase
  10.1088/1367-2630/17/5/053002} {\bibfield  {journal} {\bibinfo  {journal}
  {New Journal of Physics}\ }\textbf {\bibinfo {volume} {17}},\ \bibinfo {eid}
  {053002} (\bibinfo {year} {2015})},\ \Eprint {http://arxiv.org/abs/1408.5140}
  {arXiv:1408.5140 [quant-ph]} \BibitemShut {NoStop}%
\bibitem [{\citenamefont {Rams}\ \emph {et~al.}(2015)\citenamefont {Rams},
  \citenamefont {Zauner}, \citenamefont {Bal}, \citenamefont {Haegeman},\ and\
  \citenamefont {Verstraete}}]{Rams_2015}%
  \BibitemOpen
  \bibfield  {author} {\bibinfo {author} {\bibfnamefont {M.~M.}\ \bibnamefont
  {Rams}}, \bibinfo {author} {\bibfnamefont {V.}~\bibnamefont {Zauner}},
  \bibinfo {author} {\bibfnamefont {M.}~\bibnamefont {Bal}}, \bibinfo {author}
  {\bibfnamefont {J.}~\bibnamefont {Haegeman}}, \ and\ \bibinfo {author}
  {\bibfnamefont {F.}~\bibnamefont {Verstraete}},\ }\href {\doibase
  10.1103/physrevb.92.235150} {\bibfield  {journal} {\bibinfo  {journal}
  {Physical Review B}\ }\textbf {\bibinfo {volume} {92}} (\bibinfo {year}
  {2015}),\ 10.1103/physrevb.92.235150}\BibitemShut {NoStop}%
\bibitem [{\citenamefont {Bal}\ \emph {et~al.}(2016)\citenamefont {Bal},
  \citenamefont {Rams}, \citenamefont {Zauner}, \citenamefont {Haegeman},\ and\
  \citenamefont {Verstraete}}]{Bal_2016}%
  \BibitemOpen
  \bibfield  {author} {\bibinfo {author} {\bibfnamefont {M.}~\bibnamefont
  {Bal}}, \bibinfo {author} {\bibfnamefont {M.~M.}\ \bibnamefont {Rams}},
  \bibinfo {author} {\bibfnamefont {V.}~\bibnamefont {Zauner}}, \bibinfo
  {author} {\bibfnamefont {J.}~\bibnamefont {Haegeman}}, \ and\ \bibinfo
  {author} {\bibfnamefont {F.}~\bibnamefont {Verstraete}},\ }\href {\doibase
  10.1103/physrevb.94.205122} {\bibfield  {journal} {\bibinfo  {journal}
  {Physical Review B}\ }\textbf {\bibinfo {volume} {94}} (\bibinfo {year}
  {2016}),\ 10.1103/physrevb.94.205122}\BibitemShut {NoStop}%
\bibitem [{\citenamefont {Nishino}\ \emph {et~al.}(1996)\citenamefont
  {Nishino}, \citenamefont {Okunishi},\ and\ \citenamefont
  {Kikuchi}}]{nishino1996numerical}%
  \BibitemOpen
  \bibfield  {author} {\bibinfo {author} {\bibfnamefont {T.}~\bibnamefont
  {Nishino}}, \bibinfo {author} {\bibfnamefont {K.}~\bibnamefont {Okunishi}}, \
  and\ \bibinfo {author} {\bibfnamefont {M.}~\bibnamefont {Kikuchi}},\
  }\href@noop {} {\bibfield  {journal} {\bibinfo  {journal} {Physics Letters
  A}\ }\textbf {\bibinfo {volume} {213}},\ \bibinfo {pages} {69} (\bibinfo
  {year} {1996})}\BibitemShut {NoStop}%
\bibitem [{\citenamefont {{Pollmann}}\ \emph {et~al.}(2009)\citenamefont
  {{Pollmann}}, \citenamefont {{Mukerjee}}, \citenamefont {{Turner}},\ and\
  \citenamefont {{Moore}}}]{Pollmann_2008}%
  \BibitemOpen
  \bibfield  {author} {\bibinfo {author} {\bibfnamefont {F.}~\bibnamefont
  {{Pollmann}}}, \bibinfo {author} {\bibfnamefont {S.}~\bibnamefont
  {{Mukerjee}}}, \bibinfo {author} {\bibfnamefont {A.~M.}\ \bibnamefont
  {{Turner}}}, \ and\ \bibinfo {author} {\bibfnamefont {J.~E.}\ \bibnamefont
  {{Moore}}},\ }\href {\doibase 10.1103/PhysRevLett.102.255701} {\bibfield
  {journal} {\bibinfo  {journal} {\prl}\ }\textbf {\bibinfo {volume} {102}},\
  \bibinfo {eid} {255701} (\bibinfo {year} {2009})},\ \Eprint
  {http://arxiv.org/abs/0812.2903} {arXiv:0812.2903 [cond-mat.str-el]}
  \BibitemShut {NoStop}%
\bibitem [{\citenamefont {Tagliacozzo}\ \emph {et~al.}(2008)\citenamefont
  {Tagliacozzo}, \citenamefont {de~Oliveira}, \citenamefont {Iblisdir},\ and\
  \citenamefont {Latorre}}]{Tagliacozzo2008}%
  \BibitemOpen
  \bibfield  {author} {\bibinfo {author} {\bibfnamefont {L.}~\bibnamefont
  {Tagliacozzo}}, \bibinfo {author} {\bibfnamefont {T.~R.}\ \bibnamefont
  {de~Oliveira}}, \bibinfo {author} {\bibfnamefont {S.}~\bibnamefont
  {Iblisdir}}, \ and\ \bibinfo {author} {\bibfnamefont {J.~I.}\ \bibnamefont
  {Latorre}},\ }\href {\doibase 10.1103/PhysRevB.78.024410} {\bibfield
  {journal} {\bibinfo  {journal} {Phys. Rev. B}\ }\textbf {\bibinfo {volume}
  {78}},\ \bibinfo {pages} {024410} (\bibinfo {year} {2008})}\BibitemShut
  {NoStop}%
\bibitem [{\citenamefont {{Pirvu}}\ \emph {et~al.}(2012)\citenamefont
  {{Pirvu}}, \citenamefont {{Vidal}}, \citenamefont {{Verstraete}},\ and\
  \citenamefont {{Tagliacozzo}}}]{Pirvu_2012}%
  \BibitemOpen
  \bibfield  {author} {\bibinfo {author} {\bibfnamefont {B.}~\bibnamefont
  {{Pirvu}}}, \bibinfo {author} {\bibfnamefont {G.}~\bibnamefont {{Vidal}}},
  \bibinfo {author} {\bibfnamefont {F.}~\bibnamefont {{Verstraete}}}, \ and\
  \bibinfo {author} {\bibfnamefont {L.}~\bibnamefont {{Tagliacozzo}}},\ }\href
  {\doibase 10.1103/PhysRevB.86.075117} {\bibfield  {journal} {\bibinfo
  {journal} {\prb}\ }\textbf {\bibinfo {volume} {86}},\ \bibinfo {eid} {075117}
  (\bibinfo {year} {2012})},\ \Eprint {http://arxiv.org/abs/1204.3934}
  {arXiv:1204.3934 [cond-mat.stat-mech]} \BibitemShut {NoStop}%
\bibitem [{\citenamefont {Vanhecke}\ \emph {et~al.}(2019)\citenamefont
  {Vanhecke}, \citenamefont {Haegeman}, \citenamefont {{Van Acoleyen}},
  \citenamefont {Vanderstraeten},\ and\ \citenamefont
  {Verstraete}}]{vanhecke2019scaling}%
  \BibitemOpen
  \bibfield  {author} {\bibinfo {author} {\bibfnamefont {B.}~\bibnamefont
  {Vanhecke}}, \bibinfo {author} {\bibfnamefont {J.}~\bibnamefont {Haegeman}},
  \bibinfo {author} {\bibfnamefont {K.}~\bibnamefont {{Van Acoleyen}}},
  \bibinfo {author} {\bibfnamefont {L.}~\bibnamefont {Vanderstraeten}}, \ and\
  \bibinfo {author} {\bibfnamefont {F.}~\bibnamefont {Verstraete}},\
  }\href@noop {} {\enquote {\bibinfo {title} {A scaling hypothesis for matrix
  product states},}\ } (\bibinfo {year} {2019}),\ \Eprint
  {http://arxiv.org/abs/1907.08603} {arXiv:1907.08603 [cond-mat.stat-mech]}
  \BibitemShut {NoStop}%
\bibitem [{\citenamefont {{Vidal}}(2007)}]{Vidal_2007}%
  \BibitemOpen
  \bibfield  {author} {\bibinfo {author} {\bibfnamefont {G.}~\bibnamefont
  {{Vidal}}},\ }\href {\doibase 10.1103/PhysRevLett.99.220405} {\bibfield
  {journal} {\bibinfo  {journal} {\prl}\ }\textbf {\bibinfo {volume} {99}},\
  \bibinfo {eid} {220405} (\bibinfo {year} {2007})},\ \Eprint
  {http://arxiv.org/abs/cond-mat/0512165} {arXiv:cond-mat/0512165
  [cond-mat.str-el]} \BibitemShut {NoStop}%
\bibitem [{\citenamefont {{Vidal}}(2008)}]{vidal_class_2008}%
  \BibitemOpen
  \bibfield  {author} {\bibinfo {author} {\bibfnamefont {G.}~\bibnamefont
  {{Vidal}}},\ }\href@noop {} {\bibfield  {journal} {\bibinfo  {journal} {Phys.
  Rev. Lett.}\ }\textbf {\bibinfo {volume} {101}},\ \bibinfo {pages} {110501}
  (\bibinfo {year} {2008})},\ \Eprint {http://arxiv.org/abs/quant-ph/0610099}
  {arXiv:quant-ph/0610099} \BibitemShut {NoStop}%
\bibitem [{\citenamefont {Evenbly}\ and\ \citenamefont
  {Vidal}(2009)}]{Evenbly_2009}%
  \BibitemOpen
  \bibfield  {author} {\bibinfo {author} {\bibfnamefont {G.}~\bibnamefont
  {Evenbly}}\ and\ \bibinfo {author} {\bibfnamefont {G.}~\bibnamefont
  {Vidal}},\ }\href {\doibase 10.1103/physrevb.79.144108} {\bibfield  {journal}
  {\bibinfo  {journal} {Physical Review B}\ }\textbf {\bibinfo {volume} {79}}
  (\bibinfo {year} {2009}),\ 10.1103/physrevb.79.144108}\BibitemShut {NoStop}%
\bibitem [{\citenamefont {{Giovannetti}}\ \emph {et~al.}(2008)\citenamefont
  {{Giovannetti}}, \citenamefont {{Montangero}},\ and\ \citenamefont
  {{Fazio}}}]{Giovannetti_2008}%
  \BibitemOpen
  \bibfield  {author} {\bibinfo {author} {\bibfnamefont {V.}~\bibnamefont
  {{Giovannetti}}}, \bibinfo {author} {\bibfnamefont {S.}~\bibnamefont
  {{Montangero}}}, \ and\ \bibinfo {author} {\bibfnamefont {R.}~\bibnamefont
  {{Fazio}}},\ }\href {\doibase 10.1103/PhysRevLett.101.180503} {\bibfield
  {journal} {\bibinfo  {journal} {\prl}\ }\textbf {\bibinfo {volume} {101}},\
  \bibinfo {eid} {180503} (\bibinfo {year} {2008})},\ \Eprint
  {http://arxiv.org/abs/0804.0520} {arXiv:0804.0520 [quant-ph]} \BibitemShut
  {NoStop}%
\bibitem [{\citenamefont {Pfeifer}\ \emph {et~al.}(2009)\citenamefont
  {Pfeifer}, \citenamefont {Evenbly},\ and\ \citenamefont
  {Vidal}}]{Pfeifer_2009}%
  \BibitemOpen
  \bibfield  {author} {\bibinfo {author} {\bibfnamefont {R.~N.~C.}\
  \bibnamefont {Pfeifer}}, \bibinfo {author} {\bibfnamefont {G.}~\bibnamefont
  {Evenbly}}, \ and\ \bibinfo {author} {\bibfnamefont {G.}~\bibnamefont
  {Vidal}},\ }\href {\doibase 10.1103/physreva.79.040301} {\bibfield  {journal}
  {\bibinfo  {journal} {Physical Review A}\ }\textbf {\bibinfo {volume} {79}}
  (\bibinfo {year} {2009}),\ 10.1103/physreva.79.040301}\BibitemShut {NoStop}%
\bibitem [{\citenamefont {Evenbly}\ and\ \citenamefont
  {Vidal}(2011)}]{evenbly2011quantum}%
  \BibitemOpen
  \bibfield  {author} {\bibinfo {author} {\bibfnamefont {G.}~\bibnamefont
  {Evenbly}}\ and\ \bibinfo {author} {\bibfnamefont {G.}~\bibnamefont
  {Vidal}},\ }\href@noop {} {\enquote {\bibinfo {title} {Quantum criticality
  with the multi-scale entanglement renormalization ansatz},}\ } (\bibinfo
  {year} {2011}),\ \Eprint {http://arxiv.org/abs/1109.5334} {arXiv:1109.5334
  [quant-ph]} \BibitemShut {NoStop}%
\bibitem [{\citenamefont {Evenbly}\ and\ \citenamefont
  {Vidal}(2015)}]{Evenbly_2015}%
  \BibitemOpen
  \bibfield  {author} {\bibinfo {author} {\bibfnamefont {G.}~\bibnamefont
  {Evenbly}}\ and\ \bibinfo {author} {\bibfnamefont {G.}~\bibnamefont
  {Vidal}},\ }\href {\doibase 10.1103/physrevlett.115.200401} {\bibfield
  {journal} {\bibinfo  {journal} {Physical Review Letters}\ }\textbf {\bibinfo
  {volume} {115}} (\bibinfo {year} {2015}),\
  10.1103/physrevlett.115.200401}\BibitemShut {NoStop}%
\bibitem [{\citenamefont {Czech}\ \emph {et~al.}(2016)\citenamefont {Czech},
  \citenamefont {Evenbly}, \citenamefont {Lamprou}, \citenamefont {McCandlish},
  \citenamefont {Qi}, \citenamefont {Sully},\ and\ \citenamefont
  {Vidal}}]{czech_tensor_2016}%
  \BibitemOpen
  \bibfield  {author} {\bibinfo {author} {\bibfnamefont {B.}~\bibnamefont
  {Czech}}, \bibinfo {author} {\bibfnamefont {G.}~\bibnamefont {Evenbly}},
  \bibinfo {author} {\bibfnamefont {L.}~\bibnamefont {Lamprou}}, \bibinfo
  {author} {\bibfnamefont {S.}~\bibnamefont {McCandlish}}, \bibinfo {author}
  {\bibfnamefont {X.-l.}\ \bibnamefont {Qi}}, \bibinfo {author} {\bibfnamefont
  {J.}~\bibnamefont {Sully}}, \ and\ \bibinfo {author} {\bibfnamefont
  {G.}~\bibnamefont {Vidal}},\ }\href {\doibase 10.1103/PhysRevB.94.085101}
  {\bibfield  {journal} {\bibinfo  {journal} {Physical Review B}\ }\textbf
  {\bibinfo {volume} {94}},\ \bibinfo {pages} {085101} (\bibinfo {year}
  {2016})}\BibitemShut {NoStop}%
\bibitem [{\citenamefont {{Evenbly}}\ \emph {et~al.}(2010)\citenamefont
  {{Evenbly}}, \citenamefont {{Pfeifer}}, \citenamefont {{Pic{\'o}}},
  \citenamefont {{Iblisdir}}, \citenamefont {{Tagliacozzo}}, \citenamefont
  {{McCulloch}},\ and\ \citenamefont {{Vidal}}}]{Evenbly_2010}%
  \BibitemOpen
  \bibfield  {author} {\bibinfo {author} {\bibfnamefont {G.}~\bibnamefont
  {{Evenbly}}}, \bibinfo {author} {\bibfnamefont {R.~N.~C.}\ \bibnamefont
  {{Pfeifer}}}, \bibinfo {author} {\bibfnamefont {V.}~\bibnamefont
  {{Pic{\'o}}}}, \bibinfo {author} {\bibfnamefont {S.}~\bibnamefont
  {{Iblisdir}}}, \bibinfo {author} {\bibfnamefont {L.}~\bibnamefont
  {{Tagliacozzo}}}, \bibinfo {author} {\bibfnamefont {I.~P.}\ \bibnamefont
  {{McCulloch}}}, \ and\ \bibinfo {author} {\bibfnamefont {G.}~\bibnamefont
  {{Vidal}}},\ }\href {\doibase 10.1103/PhysRevB.82.161107} {\bibfield
  {journal} {\bibinfo  {journal} {\prb}\ }\textbf {\bibinfo {volume} {82}},\
  \bibinfo {eid} {161107} (\bibinfo {year} {2010})},\ \Eprint
  {http://arxiv.org/abs/0912.1642} {arXiv:0912.1642 [cond-mat.str-el]}
  \BibitemShut {NoStop}%
\bibitem [{\citenamefont {{Evenbly}}\ and\ \citenamefont
  {{Vidal}}(2013)}]{Evenbly_2013}%
  \BibitemOpen
  \bibfield  {author} {\bibinfo {author} {\bibfnamefont {G.}~\bibnamefont
  {{Evenbly}}}\ and\ \bibinfo {author} {\bibfnamefont {G.}~\bibnamefont
  {{Vidal}}},\ }\href@noop {} {\bibfield  {journal} {\bibinfo  {journal} {arXiv
  e-prints}\ ,\ \bibinfo {eid} {arXiv:1307.0831}} (\bibinfo {year} {2013})},\
  \Eprint {http://arxiv.org/abs/1307.0831} {arXiv:1307.0831 [quant-ph]}
  \BibitemShut {NoStop}%
\bibitem [{\citenamefont {Cardy}(1984)}]{Cardy_1984}%
  \BibitemOpen
  \bibfield  {author} {\bibinfo {author} {\bibfnamefont {J.~L.}\ \bibnamefont
  {Cardy}},\ }\href {\doibase 10.1088/0305-4470/17/7/003} {\bibfield  {journal}
  {\bibinfo  {journal} {Journal of Physics A: Mathematical and General}\
  }\textbf {\bibinfo {volume} {17}},\ \bibinfo {pages} {L385} (\bibinfo {year}
  {1984})}\BibitemShut {NoStop}%
\bibitem [{\citenamefont {Cardy}(1986)}]{Cardy:1986ie}%
  \BibitemOpen
  \bibfield  {author} {\bibinfo {author} {\bibfnamefont {J.~L.}\ \bibnamefont
  {Cardy}},\ }\href {\doibase 10.1016/0550-3213(86)90552-3} {\bibfield
  {journal} {\bibinfo  {journal} {Nucl. Phys.}\ }\textbf {\bibinfo {volume}
  {B270}},\ \bibinfo {pages} {186} (\bibinfo {year} {1986})}\BibitemShut
  {NoStop}%
\bibitem [{\citenamefont {Cardy}(1988)}]{Cardy1988}%
  \BibitemOpen
  \bibfield  {author} {\bibinfo {author} {\bibfnamefont {J.}~\bibnamefont
  {Cardy}},\ }\href@noop {} {\emph {\bibinfo {title} {{Finite-size scaling}}}}\
  (\bibinfo  {publisher} {Elsevier},\ \bibinfo {year} {1988})\BibitemShut
  {NoStop}%
\bibitem [{\citenamefont {Witten}(2018)}]{Witten_2018}%
  \BibitemOpen
  \bibfield  {author} {\bibinfo {author} {\bibfnamefont {E.}~\bibnamefont
  {Witten}},\ }\href {\doibase 10.1103/revmodphys.90.045003} {\bibfield
  {journal} {\bibinfo  {journal} {Reviews of Modern Physics}\ }\textbf
  {\bibinfo {volume} {90}} (\bibinfo {year} {2018}),\
  10.1103/revmodphys.90.045003}\BibitemShut {NoStop}%
\bibitem [{\citenamefont {Di~Francesco}\ \emph {et~al.}(1997)\citenamefont
  {Di~Francesco}, \citenamefont {Mathieu},\ and\ \citenamefont
  {S\'{e}n\'{e}chal}}]{DiFrancesco:639405}%
  \BibitemOpen
  \bibfield  {author} {\bibinfo {author} {\bibfnamefont {P.}~\bibnamefont
  {Di~Francesco}}, \bibinfo {author} {\bibfnamefont {P.}~\bibnamefont
  {Mathieu}}, \ and\ \bibinfo {author} {\bibfnamefont {D.}~\bibnamefont
  {S\'{e}n\'{e}chal}},\ }\href {\doibase 10.1007/978-1-4612-2256-9} {\emph
  {\bibinfo {title} {{Conformal field theory}}}},\ Graduate texts in
  contemporary physics\ (\bibinfo  {publisher} {Springer},\ \bibinfo {address}
  {New York, NY},\ \bibinfo {year} {1997})\BibitemShut {NoStop}%
\bibitem [{\citenamefont {{Singh}}\ \emph {et~al.}(2011)\citenamefont
  {{Singh}}, \citenamefont {{Pfeifer}},\ and\ \citenamefont
  {{Vidal}}}]{singh_tensor_2011}%
  \BibitemOpen
  \bibfield  {author} {\bibinfo {author} {\bibfnamefont {S.}~\bibnamefont
  {{Singh}}}, \bibinfo {author} {\bibfnamefont {R.~N.~C.}\ \bibnamefont
  {{Pfeifer}}}, \ and\ \bibinfo {author} {\bibfnamefont {G.}~\bibnamefont
  {{Vidal}}},\ }\href {http://arxiv.org/abs/1008.4774} {\bibfield  {journal}
  {\bibinfo  {journal} {Phys. Rev. {B}}\ }\textbf {\bibinfo {volume} {83}},\
  \bibinfo {pages} {115125} (\bibinfo {year} {2011})},\ \Eprint
  {http://arxiv.org/abs/1008.4774} {arXiv:1008.4774} \BibitemShut {NoStop}%
\bibitem [{\citenamefont {{Singh}}\ and\ \citenamefont
  {{Vidal}}(2013)}]{singh_symmetry_2013}%
  \BibitemOpen
  \bibfield  {author} {\bibinfo {author} {\bibfnamefont {S.}~\bibnamefont
  {{Singh}}}\ and\ \bibinfo {author} {\bibfnamefont {G.}~\bibnamefont
  {{Vidal}}},\ }\href {http://arxiv.org/abs/1303.6716} {\bibfield  {journal}
  {\bibinfo  {journal} {Phys. Rev. {B}}\ }\textbf {\bibinfo {volume} {88}}
  (\bibinfo {year} {2013})},\ \Eprint {http://arxiv.org/abs/1303.6716}
  {arXiv:1303.6716} \BibitemShut {NoStop}%
\bibitem [{\citenamefont {Haegeman}\ \emph {et~al.}(2012)\citenamefont
  {Haegeman}, \citenamefont {Pirvu}, \citenamefont {Weir}, \citenamefont
  {Cirac}, \citenamefont {Osborne}, \citenamefont {Verschelde},\ and\
  \citenamefont {Verstraete}}]{Haegeman_2012}%
  \BibitemOpen
  \bibfield  {author} {\bibinfo {author} {\bibfnamefont {J.}~\bibnamefont
  {Haegeman}}, \bibinfo {author} {\bibfnamefont {B.}~\bibnamefont {Pirvu}},
  \bibinfo {author} {\bibfnamefont {D.~J.}\ \bibnamefont {Weir}}, \bibinfo
  {author} {\bibfnamefont {J.~I.}\ \bibnamefont {Cirac}}, \bibinfo {author}
  {\bibfnamefont {T.~J.}\ \bibnamefont {Osborne}}, \bibinfo {author}
  {\bibfnamefont {H.}~\bibnamefont {Verschelde}}, \ and\ \bibinfo {author}
  {\bibfnamefont {F.}~\bibnamefont {Verstraete}},\ }\href {\doibase
  10.1103/physrevb.85.100408} {\bibfield  {journal} {\bibinfo  {journal}
  {Physical Review B}\ }\textbf {\bibinfo {volume} {85}} (\bibinfo {year}
  {2012}),\ 10.1103/physrevb.85.100408}\BibitemShut {NoStop}%
\bibitem [{Note1()}]{Note1}%
  \BibitemOpen
  \bibinfo {note} {In our simulations, depending on the initial random seed in
  the numerical optimization we end up with isometries giving rise to two
  different types of dispersion relation: of the ferro-magnetic type fig.2a or
  of the anti-ferro-magnetic type fig.2b, with only a positive scaling operator
  $S$ in the former case. By applying a $\protect \mathbb {Z}_2$ transformation
  on the upper leg of the isometry we can switch between both cases. Using the
  $\protect \mathbb {Z}_2$ symmetry of the tensors, one can show that this is a
  MERA gauge transformation, i.e. a transformation on the MERA tensors that
  leaves the actual physical state invariant. From the same $\protect \mathbb
  {Z}_2$ symmetry of the tensors one can also show that this transformation
  shifts the momentum of the MPO eigenstates: $k\rightarrow k+m\pi $, with
  $m=(0,1)$ the $\protect \mathbb {Z}_2$ quantum number of the state. This then
  explains the $k>\pi /2$ part of the second case fig.2b as arising from the
  $m=1$ negative momentum branch of the first case. This generalises to more
  general group symmetries: contracting an element of the group $U_g$ - with
  $g$ in the center of the group - on the upper leg of the symmetric isometry,
  shifts the momentum of the MPO eigenstates accordingly, while leaving the
  physical MERA state invariant.}\BibitemShut {Stop}%
\bibitem [{\citenamefont {{Henkel}}(1987)}]{Henkel_1987}%
  \BibitemOpen
  \bibfield  {author} {\bibinfo {author} {\bibfnamefont {M.}~\bibnamefont
  {{Henkel}}},\ }\href {\doibase 10.1088/0305-4470/20/4/033} {\bibfield
  {journal} {\bibinfo  {journal} {Journal of Physics A Mathematical General}\
  }\textbf {\bibinfo {volume} {20}},\ \bibinfo {pages} {995} (\bibinfo {year}
  {1987})}\BibitemShut {NoStop}%
\bibitem [{\citenamefont {{Reinicke}}(1987)}]{Reinicke_1987}%
  \BibitemOpen
  \bibfield  {author} {\bibinfo {author} {\bibfnamefont {P.}~\bibnamefont
  {{Reinicke}}},\ }\href {\doibase 10.1088/0305-4470/20/13/048} {\bibfield
  {journal} {\bibinfo  {journal} {Journal of Physics A Mathematical General}\
  }\textbf {\bibinfo {volume} {20}},\ \bibinfo {pages} {4501} (\bibinfo {year}
  {1987})}\BibitemShut {NoStop}%
\bibitem [{\citenamefont {Giamarchi}(2004)}]{Giamarchi}%
  \BibitemOpen
  \bibfield  {author} {\bibinfo {author} {\bibfnamefont {T.}~\bibnamefont
  {Giamarchi}},\ }\href {\doibase 10.1093/acprof:oso/9780198525004.001.0001}
  {\emph {\bibinfo {title} {{Quantum physics in one dimension}}}},\ Internat.
  Ser. Mono. Phys.\ (\bibinfo  {publisher} {Clarendon Press},\ \bibinfo
  {address} {Oxford},\ \bibinfo {year} {2004})\BibitemShut {NoStop}%
\bibitem [{\citenamefont {{Hastings}}(2006)}]{Hastings_2006}%
  \BibitemOpen
  \bibfield  {author} {\bibinfo {author} {\bibfnamefont {M.~B.}\ \bibnamefont
  {{Hastings}}},\ }\href {\doibase 10.1103/PhysRevB.73.085115} {\bibfield
  {journal} {\bibinfo  {journal} {\prb}\ }\textbf {\bibinfo {volume} {73}},\
  \bibinfo {eid} {085115} (\bibinfo {year} {2006})},\ \Eprint
  {http://arxiv.org/abs/cond-mat/0508554} {arXiv:cond-mat/0508554
  [cond-mat.str-el]} \BibitemShut {NoStop}%
\bibitem [{\citenamefont {Molnar}\ \emph {et~al.}(2015)\citenamefont {Molnar},
  \citenamefont {Schuch}, \citenamefont {Verstraete},\ and\ \citenamefont
  {Cirac}}]{Molnar_2015}%
  \BibitemOpen
  \bibfield  {author} {\bibinfo {author} {\bibfnamefont {A.}~\bibnamefont
  {Molnar}}, \bibinfo {author} {\bibfnamefont {N.}~\bibnamefont {Schuch}},
  \bibinfo {author} {\bibfnamefont {F.}~\bibnamefont {Verstraete}}, \ and\
  \bibinfo {author} {\bibfnamefont {J.~I.}\ \bibnamefont {Cirac}},\ }\href
  {\doibase 10.1103/physrevb.91.045138} {\bibfield  {journal} {\bibinfo
  {journal} {Physical Review B}\ }\textbf {\bibinfo {volume} {91}} (\bibinfo
  {year} {2015}),\ 10.1103/physrevb.91.045138}\BibitemShut {NoStop}%
\bibitem [{\citenamefont {{Verstraete}}\ \emph {et~al.}(2004)\citenamefont
  {{Verstraete}}, \citenamefont {{Garc{\'\i}a-Ripoll}},\ and\ \citenamefont
  {{Cirac}}}]{Verstraete_2004MPO}%
  \BibitemOpen
  \bibfield  {author} {\bibinfo {author} {\bibfnamefont {F.}~\bibnamefont
  {{Verstraete}}}, \bibinfo {author} {\bibfnamefont {J.~J.}\ \bibnamefont
  {{Garc{\'\i}a-Ripoll}}}, \ and\ \bibinfo {author} {\bibfnamefont {J.~I.}\
  \bibnamefont {{Cirac}}},\ }\href {\doibase 10.1103/PhysRevLett.93.207204}
  {\bibfield  {journal} {\bibinfo  {journal} {\prl}\ }\textbf {\bibinfo
  {volume} {93}},\ \bibinfo {eid} {207204} (\bibinfo {year} {2004})},\ \Eprint
  {http://arxiv.org/abs/cond-mat/0406426} {arXiv:cond-mat/0406426
  [cond-mat.other]} \BibitemShut {NoStop}%
\bibitem [{\citenamefont {Evenbly}\ \emph {et~al.}(2010)\citenamefont
  {Evenbly}, \citenamefont {Corboz},\ and\ \citenamefont
  {Vidal}}]{Evenbly_2010nonlocal}%
  \BibitemOpen
  \bibfield  {author} {\bibinfo {author} {\bibfnamefont {G.}~\bibnamefont
  {Evenbly}}, \bibinfo {author} {\bibfnamefont {P.}~\bibnamefont {Corboz}}, \
  and\ \bibinfo {author} {\bibfnamefont {G.}~\bibnamefont {Vidal}},\ }\href
  {\doibase 10.1103/PhysRevB.82.132411} {\bibfield  {journal} {\bibinfo
  {journal} {Phys. Rev. B}\ }\textbf {\bibinfo {volume} {82}},\ \bibinfo
  {pages} {132411} (\bibinfo {year} {2010})}\BibitemShut {NoStop}%
\bibitem [{Note2()}]{Note2}%
  \BibitemOpen
  \bibinfo {note} {In preparation}\BibitemShut {NoStop}%
\end{thebibliography}%

\end{document}